\documentclass[twocolumn,showkeys,preprintnumbers,amsmath,amssymb,prb]{revtex4}
\usepackage{graphicx,calc}% Include figure files
\usepackage{bm}% bold math
\usepackage{enumerate}
\begin{document}
%\draft
\title{Electronic Structure and Lattice dynamics of NaFeAs}
\author{Shuiquan Deng, J\"urgen K\"ohler and Arndt Simon}
\affiliation{Max-Planck Institut f\"ur Festk\"orperforschung, D-70569 Stuttgart, Germany}
\email{s.deng@fkf.mpg.de}
%\date{\today}

\begin{abstract}
The similarity of the electronic structures of NaFeAs and other Fe pnictides has been demonstrated on the basis of first-principle calculations. The global double-degeneracy of electronic bands along X-M and R-A direction indicates the instability of Fe pnictides and is explained on the basis of a tight-binding model. The de Haas-van Alphen parameters for the Fermi surface (FS) of NaFeAs have been calculated. A $\mathbf{Q}_{M}=(1/2,1/2,0)$ spin density wave (SDW) instead of a charge density wave (CDW) ground state is predicted based on the calculated generalized susceptibility $\chi(\mathbf{q})$ and a criterion derived from a restricted Hatree-Fock model. The strongest electron-phonon (e-p) coupling has been found to involve only As, Na z-direction vibration with linear-response calculations. A possible enhancement mechanism for e-p coupling due to correlation is suggested.    
\end{abstract}

%\pacs {73.22.-f, 61.46.+w, 73.63.Nm}
\maketitle

\section{INTRODUCTION}
  The recent discovery of superconductivity in LaFeAs[O$_{1-x}$F$_{x}$] \cite{r1} with the remarkable
 critical temperature $T_{c}$ of 26 K has stimulated intense experimental and theoretical
 research activities towards exploring new superconducting systems and understanding the relevant
physics. So far, four systems represented by RFe(As,P)O (R=La,Ce,Pr,Nd),(1:1:1:1), (Ba,Sr,Ca)Fe$_{2}$As$_{2}$,(1:2:2), 
(Li,Na)FeAs (1:1:1) and Fe$_{x}$(Te,Se,S) (1:1) have been discovered. They all share the similar structural character of two- dimensional Fe-As layers formed by face sharing FeAs$_{4}$ tetrahedra sandwiched by ionic layers such as RO$_{1-x}$F$_{x}$, Ba$_{1-x}$K$_{x}$, Li$_{1-x}$,Na$_{x}$, or by van der Waals gaps in the (1:1:1:1), (1:2:2), (1:1:1), (1:1) systems, respectively. The stoichiometric systems can be doped with electrons or holes by appropriate substitutions in the ionic layers. Without doping most of the parent systems show universal spin density wave (SDW)-like long range magnetic order and a tetragonal to orthorhombic/monoclinic structural phase transition at low temperature. \cite{r2} The doping supresses both the magnetic and structural phase transitions and leads to superconductivity. The (1:1:1) system seems to be an exception, in which no phase transition was observed in early experiments.  \cite{r3,r4,r5,r6} 
\paragraph*{}
Density functional studies \cite{r7,r8,r9,r10,r11} have shown that the electronic states around the Fermi level are of predominant Fe-$3d$ character. The minor splitting of these $d$ bands is mainly due to the weak tetrahedral ligand field of As. The universal feature of nesting conditions for parts of FS is generally thought to be the origin of the magnetic and structural phase transitions. The missing transitions in the (1:1:1) system call for an explanation. So far metric details of the FS for all systems have not been addressed, which would be important to provide quantitative tests to the present understanding of Fe-based superconductors when compared with dHvA data. Furthermore, the role of phonons remains important even though earlier calculations \cite{r12,r13} which show a rather weak electron-phonon (e-p) coupling, as the importance of magnetophonon coupling is revealed in a recent calculation.\cite{r10} In this work, we investigate the electronic structure and lattice dynamics of NaFeAs, because of the aforementioned behavior and its simple structure.

\section{COMPUTATIONAL DETAILS}
For the numerical calculations the experimental crystallographic data of NaFeAs at 2.5K \cite{r5} were used. The magnetic ground state is not chosen as the starting point, because we seek a comparison with the electronic structures calculated for other Fe-As based systems. The full-potential LMTO method \cite{r14,r15} based on the density functional theory in the form of LDA+GGA \cite{r16,r17} is used for all calculations. The muffin-tin (MT) radii of 2.249,2.340 and 3.271 a.u. for Fe, As, Na, respectively, are calculated according to the maximum of overlapping Hartree potentials of neighboring atoms. \cite{r18}  The analysis of states is based on these MT radii. A 2-$\kappa$ basis set with tail energies of -1.10, -0.1 Ry, respectively, was used for the Fe-$4s$, -$4p$, -$3d$, As-$4s$,-$4p$,-$4d$ and Na-$3s$, -$3p$, -$3d$ states. While the Fe-$3p$, As-$3d$ and Na-$2s,2p$ states were treated as semi-core states, all the others were handed as core states. The charge densities and the potentials inside the MTs were expanded into harmonics up to $l_{max}$=6, while those inside the interstitial region were expanded into 32646 plane waves corresponding to a cutoff energy of 190.22Ry. A \textbf{k}-mesh of 32$\times$32$\times$28 was used to calculate the electronic structure. As the calculation of the generalized electronic susceptibility $\chi(\mathbf{q},\omega)$ is sensitive to the \textbf{k}-mesh, we have chosen a mesh of 44$\times$44$\times$40, which is similar to that used in Ref.\cite{r7}. For the linear-response calculations, a coarse mesh of 12$\times$12$\times$12 and a dense mesh of 36$\times$36$\times$36 were used. The latter was used to attain the accurate integration weights in the linear-response calculations. A \textbf{q}-mesh of 6$\times$6$\times$4 was adopted to produce 30 independent phonons. Except for these parameters, all of the other parameters are the same as those in the electronic structure calculations.

\section{Results and discussion}	
\begin{figure}
\includegraphics[scale=1.0]{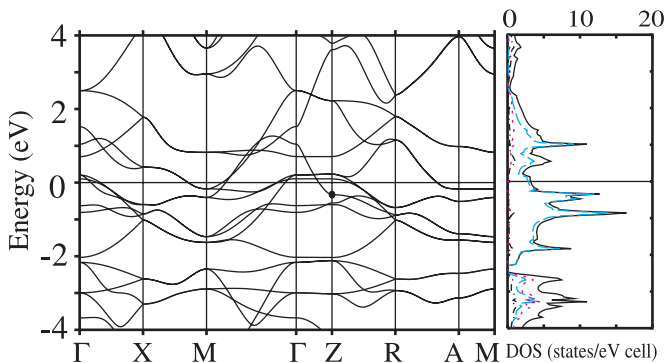}
\caption{Band structure calculated from FP-LMTO method, the black dot indicating a representative state, $\psi_{Z,12}$, for the steep band. The curves in the right part of the figure are solid for total DOS, blue broken line for Fe-$3d$, red broken line for As-$4p$, and black broken line for Na.} \label{Fig1}
\end{figure} 
\subsection{Electronic structure: General features}
The calculated FP-LMTO band structure and the electronic DOS for NaFeAs are shown in Fig. \ref{Fig1}. It is clear from the projected DOS curves that within the energy window of $E_{f}\pm 2 $ eV, the electronic states
are of predominant Fe-$3d$ character which determines the physical properties of the system. By using the TB-LMTO method \cite{r19}, we have calculated the hopping integrals defined as
\begin{equation}
    H_{ij}=\langle i|H|j\rangle ,
\end{equation}
where $\vert i \rangle$ indicates a orbital at site $i$ and $H$ is the LDA Hamiltonian. The three most  important hopping integrals for adjacent Fe ions (2.791\AA{}) are $H_{d_{xy}-d_{xy}}$=-0.284, $H_{d_{x^{2}-y^{2}}-d_{x^{2}-y^{2}}}$=0.230, and $H_{d_{xy}-d_{3z^{2}-1}}$=-0.139 eV, while the most significant hopping integral for the next nearest Fe neighbors (3.947\AA{}) is $H_{d_{x^{2}-y^{2}}-d_{x^{2}-y^{2}}}$=-0.036 eV.
Below a small energy gap the hybridization region of covalent Fe-As and partially covalent Na-As bonding sets in. The calculated dominant hopping integrals between As and Fe (2.428\AA{}) are $H_{p_{y}-d_{yz}}$=0.816, $H_{p_{z}-d_{x^{2}-y^{2}}}$=-0.620, $H_{p_{z}-d_{3z^{2}-1}}$=-0.428, $H_{p_{y}-d_{3z^{2}-1}}$=0.427 eV. Though the hopping integral between Na-$s$ and As-$s,p$ orbitals are quite large (e.g. $H_{s-s}$=1.256 eV, $H_{s-s}$=1.218 eV for the Na-As distances 2.984 and 3.107\AA{}, respectively), the rather small projected DOS of Na in the energy range of interest leads to a much weaker Na-As covalent bonding as compared to Fe-As and Fe-Fe. The strong Fe-As covalency determines the bonding in the 2D Fe-As layer, while the partial ionic bonding between Na-As determines the packing of these layers. It in important to emphasize that the covalency between Na and As is neglected in many studies, due to the choice of a smaller MT radius for Na, which assigns some electronic states to interstitial electronic states. 
\paragraph*{}
The electronic structure as shown in Fig. \ref{Fig1} is very similar to that of LiFeAs  
calculated by other methods. \cite{r9,r20} Though the detailed comparison of electronic structures of different systems is difficult due to limited data and different choices of symmetry directions, the overall agreement of the energy spectrum around the Fermi level is obvious. 
\subsection{Electronic Structure: Double Degeneracy}
An interesting feature which has not received much attention is the double degeneracy of all bands along X-M and R-A directions. Nekrasov \textit{et al.} \cite{r21} have mentioned that there are two degenerate bands along the X-M direction in the (1:1:1:1) system, however, without any further discussion. As the symmetry of the X-M line is identical to that of the R-A line, we here concentrate on the X-M line. The symmetry groups at X and M points are $D_{2h}$ and $D_{4h}$, respectively. The $D_{2h}$ symmetry at X and the $C_{2v}$ symmetry of points along X-M do not enforce any degeneracy of the bands, since $D_{2h}$ and $C_{2v}$ have only 1D representations. The $D_{4h}$ symmetry at M, however, results in double degeneracy for bands belonging to the $E_{g}$ or $E_{u}$ representations. From M to X, this degeneracy will generally be reduced due to symmetry lowering, e.g. $E_{g}\rightarrow A_{2}+B_{2}$. Obviously, the global double degeneracy of bands along the X-M line is not caused by the known symmetry. Considering the global property of this degeneracy, a hidden symmetry needs to be investigated.
\paragraph*{}
The predominant Fe-$3d$ character of all bands within $E_{f}\pm 2 $ eV indicates that these bands originate from pure Fe lattice. 
\begin{figure}
\includegraphics[scale=0.8]{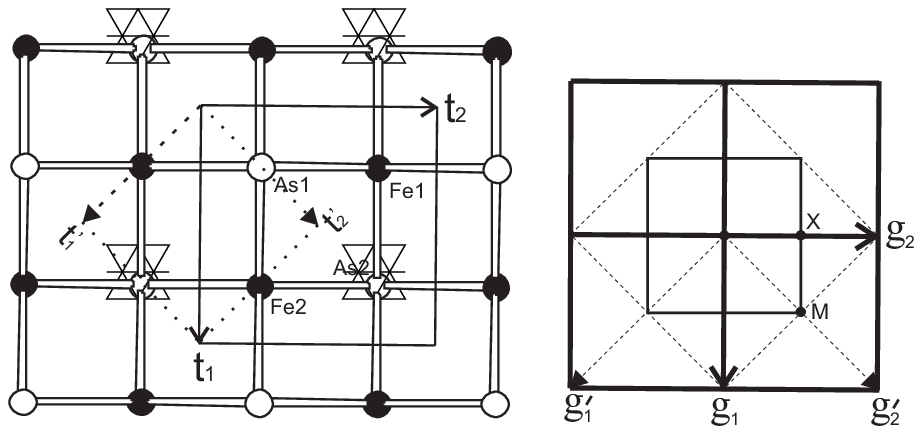}
\caption{The left figure shows the structure of the Fe-As layer in NaFeAs with $\mathbf{t}_{i}$, i=1,2 representing the primitive lattice vectors of the crystallographic unit cell; $\mathbf{t}_{i}'$ are the primitive vectors implied by the electronic band structure. The right figure shows the corresponding reciprocal lattice vectors and BZs.}\label{Fig2}
\end{figure}
It is obvious from Fig. \ref{Fig2} that each of the two As square lattices projects onto the center of the other lattice, each of which is crystallographically equivalent to the Fe lattice. Therefore a tight-binding energy dispersion for a Fe-$d$ band can be derived from a $2\times 2$ secular matrix as follows,
\begin{displaymath}
\left( \begin{array}{cc}
 H_{11}-\varepsilon & H_{12}\\
 H_{21}& H_{22}-\varepsilon
\end{array}\right)=0 ,
\end{displaymath}
where the indices 1,2 indicate Fe1 and Fe2 atoms in the unit cell, respectively. The on-site energy $H_{11}=H_{22}$ of the chosen orbital is calculated up to the second nearest neighbors, and is given for the X-M line, by
\begin{equation}
    H_{11}=\alpha_{d}+\beta_{1}^{d}\left(2\cos(2 \pi x)-2\right) -4\beta_{2}^{d}\cos(2\pi x) ,
\end{equation}
where $\alpha_{d},\beta_{1}^{d},\beta_{2}^{d}$ are the on-site energy, nearest neighbor, and second nearest neighbor hopping integrals between Fe1-d orbitals, respectively. The parameter $x\in [0,1/2]$ specifies a $\mathbf{k}$ point along the X-M line with $\mathbf{k}=x \mathbf{g}_{1}+ 1/2 \mathbf{g}_{2}$, where $\mathbf{g}_{1}$ and $\mathbf{g}_{2}$ are the reciprocal lattice vectors corresponding to $\mathbf{t}_{1}$ and $\mathbf{t}_{2}$, respectively. An interesting result is that $H_{12}=H_{21}^{*}=0$ for $\mathbf{k}$ points on the X-M line. In the tight-binding approximation the conditions of $H_{11}=H_{22}$ and $H_{12}=H_{21}^{*}=0$ result in a doubly degenerate band dispersion along the X-M line, i.e. $E(k)=H_{11}$. 
\paragraph*{}
In fact, if only Fe bands are concerned, a smaller unit cell as shown in Fig. \ref{Fig2} for the Fe sub-lattice, with the unit cell vectors, $\mathbf{t}_{1}^{'}=1/2(\mathbf{t}_{1}-t_{2})$, $\mathbf{t}_{2}^{'}=1/2(\mathbf{t}_{1}+\mathbf{t}_{2})$, $\mathbf{t}_{3}^{'}=\mathbf{t}_{3}$ can be found, which results in a double sized first BZ with the X-M line as an internal line. When the reduced zone scheme is used to represent the bands, the bands in the large BZ will be folded into the smaller BZ as shown in Fig. \ref{Fig2} by the reciprocal lattice vector. Accordingly the degeneracy of the boundary states will be doubled, as can be easily understood from the one-dimensional superstructure case. \cite{r22,r23}    
\paragraph*{}

It is obvious that the band-folding explanation can not be directly applied to the As bands or to the Fe, As hybridized bands, since the As sublattice can not be described by $\mathbf{t}_{i}^{'},i=1,2,3$. The two As atoms in the unit cell are connected by a glide plane operation. To understand the double degeneracy of these bands along X-M, we analyze the Hamiltonian matrix constructed by tight-binding approximation. We do not attempt to solve the whole secular problem which is a $28\times28$ matrix, if all Na-$3s$, As-$4s,4p$, and Fe-$4s,4p,3d$ orbitals are considered. Instead we investigate the most representative and important subblock by exploiting the symmetry. Guided by the electronic structure shown in Fig. \ref{Fig1}, we consider in the following the As-$p$ and Fe-$d$ orbitals only. By accounting for the $C_{2v}$ symmetry of the X-M line, we choose the $B_{2}$ subblock as an example. As only $p_{z}$ and $d_{xz}$ belong to the $B_{2}$ irreducible representation, we attain a $4\times 4$ matrix secular problem:
\begin{displaymath}
\left( \begin{array}{llll}
 H_{11}-\varepsilon&0 &0 &H_{41}^{*}\\
 0 & H_{11}-\varepsilon &H_{32}^{*} &0\\
0 &H_{32} &H_{33}-\varepsilon &0\\
H_{41} &0 &0 &H_{33}-\varepsilon 
\end{array}\right)=0,
\end{displaymath}
where the Hamiltonian matrix elements in both the rows and columns are arranged in the order of Fe1, Fe2, As1 and As2 (Fig. \ref{Fig2}), respectively. Obviously the upper $2\times 2$ diagonal subblock corresponds to the Fe sublattice, while the lower diagonal subblock corresponds to the As sublattice. Though the As sublattice has a different symmetry as compared to the Fe sublattice, their tight-binding Hamiltonian matrix have the same diagonal structure. Our calculations based on the structure as shown in Fig. \ref{Fig2} indicate that $H_{43}=H_{34}^{*}=0$, while $H_{33}=H_{44}$ calculated up to the 2nd nearest neighbors of As is given by
\begin{equation}
    H_{33}=\alpha_{p}+\beta_{1}^{p} \left(2 \cos(2 \pi x)-2\right) -4\beta_{2}^{p}\cos(2\pi x),
\end{equation} 
where $\alpha_{p},\beta_{1}^{p},\beta_{2}^{p}$ are the onsite energy, nearest neighbor, second nearest neighbor hopping integrals between As1-$p_{z}$ orbitals, respectively. The similarity between the Fe, As subblocks of the secular matrix suggests that the bands of the As sublattice should also be doubly degenerate along the X-M line. When the Fe-As interactions are taken into account, the situation becomes more complicated, because there are non-zero off-diagonal matrix elements. Under the aforementioned approximations, the matrix elements $H_{31}=H_{13}^{*}=0$ and $H_{42}=H_{24}^{*}=0$ are obtained. $H_{32}$ is found to be equal to the complex conjugate of $H_{41}$, i.e. $H_{32}$=$H_{41}$, and is given by
\begin{equation}
    H_{32}=\beta_{pd}\left( 1+e^{-i 2 \pi x}\right) + \beta_1^{pd}\left(-2-2 e^{-i 2 \pi x}\right),
\end{equation}
where $\beta_{pd}$, $\beta_{1}^{pd}$, are the nearest neighbor, second nearest neighbor hopping integrals between As1-$p_{z}$ and Fe2-$d_{xz}$ orbital , respectively. Simple algebraic operations of the $4\times4$ secular matrix equation lead to the following reduced problem:
\begin{eqnarray}
&\left((H_{11}-\varepsilon)(H_{33}-\varepsilon)-\arrowvert H_{32}\arrowvert ^{2} \right) \nonumber \\
&\times \left((H_{11}-\varepsilon)(H_{33}-\varepsilon)-\arrowvert H_{41}\arrowvert ^{2} \right)=0. \label{eq5}
\end{eqnarray}
Since $H_{32}=H_{41}^{*}$, Eq. \ref{eq5} which is of fourth order is reduced to two identical equations, resulting in two identical sets of roots, namely,
\begin{equation}
    \varepsilon=\frac{H_{11}+H_{33}}{2}\pm \frac{1}{2} \sqrt{(H_{11}-H_{33})^{2}+4 \arrowvert H_{32}\arrowvert ^{2}}. \label{eq6}
\end{equation} 
This result provides an explanation for the double degeneracy of bands along X-M line. As can be seen from Eq. \ref{eq6} if the coupling between Fe and As sublattices is removed, i.e. $H_{32}=0$, the situations of pure Fe and As lattices are recovered. Within our tight-binding model, the occurrence of zero matrix elements and their positions are essential to the double degeneracy. Our calculations show that this is a combined result for the site symmetry of Fe, As and the special direction X-M, which makes the terms $e^{in_{1}2\pi x} e^{in_{2}\pi}H_{ij}$, $e^{in_{1}2\pi x} e^{in_{2}'\pi}H_{ij}$ with $\vert n_{2}-n_{2}' \vert = odd $, always occur in pairs and thus cancels with each other, where $n_{1} \mathbf{t}_{1}+n_{2} \mathbf{t}_{2}$ indicates the position of the target atom for the hopping integral of interest. This conclusion holds even when the interlayer hopping terms are included, because $\mathbf{t}_{3}$ is perpendicular to $\mathbf{k}$ along the X-M and R-A directions, and the algebraic structure of the secular matrix remains unchanged. Therefore, for the bands along X-M or R-A the system can be described with the smaller unit cell as shown in Fig. \ref{Fig2}. Granted other conditions remain unchanged, the reduction of a unit cell implies a higher symmetry in the sense that more translational operations are allowed. The global degeneracy of bands along X-M and R-A directions simply infers an instability of the system. 
\begin{figure}
\includegraphics[scale=0.8]{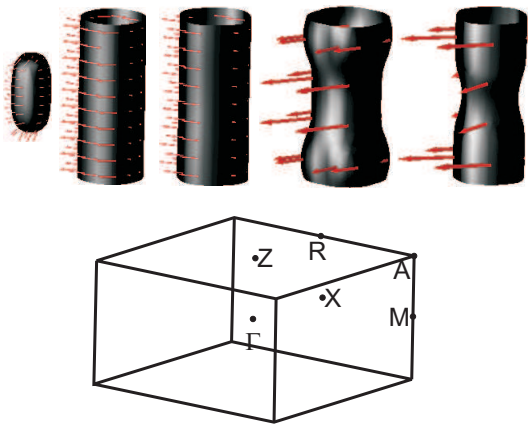}
\caption{The calculated FS from FP-LMTO method are shown on the top, the first three objects of hole nature center at $\Gamma$ with $\Gamma$-Z as their center axis. The last two objects of electronic nature center at M with M-A as their center axis. The arrows on the surface show both the magnitude and direction of the velocity field mapped onto the FS. The first BZ used throughout this work is shown below the FS.}\label{Fig3}
\end{figure}
\subsection{Electronic Structure:  Properties of Fermi Surface}
The calculated FS shown in Fig. \ref{Fig3} is composed of a small ovoid ball, two large tubes with hole character around $\Gamma$ and the other two tubes with electron character around M, respectively. Qualitatively these features agree quite well with those of other systems.\cite{r7,r9,r20} A simple visual inspection reveals possible Fermi surface nesting between the hole tubes around $\Gamma$ and electron tubes around M corresponding to a vector of (1/2,1/2,0). The calculated group velocities of electrons/holes are mapped onto the Fermi surface as shown in Fig. \ref{Fig3}, where the group velocity is defined by,
\begin{equation}
 \mathbf{v}(\mathbf{k}j)=\dfrac{1}{\hbar}\nabla E(\mathbf{k}j).
\end{equation}
The calculated Fermi velocities of electrons are much larger than those of the holes (see Fig. \ref{Fig3}). The directions of the velocities suggest that the movement of most electrons and holes are confined to the Fe-As layers. Only a few holes which are located at both ends of the ovoid ball move perpendicular to the layers. This can also be seen in Fig. \ref{Fig1}, where a steep band crosses the Fermi level along the $\Gamma$-Z direction. The property of this band will be discussed later. So far, there have been no experimental result to prove the existence and quantitative details of the FS for NaFeAs and also for other systems reliable experimental data have not been established. Calculations emphasizing the correlation effect \cite{r24} predict a large scattering rate and short life time for states around the Fermi level,  and thus disappearance of hole pockets at $\Gamma$ even in the case of a doped (1:1:1:1) system. The issue here is whether LDA has considered in a sufficient way the correlation effects. For a further test of different approaches, we have also calculated the metric details of the Fermi surface as summarized in Table \ref{tbl1}. 
\begin{table}
\caption{Calculated Metric Details of Fermi Surface and Band Masses} \label{tbl1}
\begin{ruledtabular}
\begin{tabular}{ccc}
$Orbit$\footnotemark[1] & $\mathit{F}_{calc}$(Tesla)  &$\mathit{m}_{calc}$ \\
\hline
$_{12}\Gamma_{z}^{h}$ &627.3 &-.473\\
$_{12}\Gamma_{y}^{h}$ &1452.2 &-.776\\
$_{13}\Gamma_{z}^{h}$ &1692.9 &-1.940\\
$_{13}\mathit{Z}_{z}^{h}$ &1666.0 &-1.954\\
$_{14}\Gamma_{z}^{h}$ &2078.6 &-1.480\\
$_{14}\mathit{Z}_{z}^{h}$ &2155.4 &-1.415\\
$_{15}\mathit{M}_{z}^{e}$ &1836.7 &0.918\\
$_{15}\mathit{A}_{z}^{e}$ &2820.9 &1.673\\
$_{16}\mathit{M}_{z}^{e}$ &1392.2 &1.108\\
$_{16}\mathit{A}_{z}^{e}$ &2116.3 &0.810\\
\end{tabular}
\end{ruledtabular}
\footnotetext[1] {Orbit is symbolized with $_{band\,  index}\mathit{k}_{normal}^{orbital\,type}$, where $k$ indicates the center of the orbit; $\mathit{F}_{calc}$ indicates the calculated area of an orbit with specified center and normal;Band masses are given in unit of free-electron mass.}
\end{table}
The band masses shown in Table \ref{tbl1} are calculated as $m_{calc}=\hbar/2 \pi\,dA/dE$. \cite{r25,r26} It needs to be pointed out that the calculated Fermi surface may split due to a SDW of the actual ground state. However, as this split is not too large, the calculated area of various orbits may still be valid to be compared with the de Hass-van Alphen experiments where available. 
\subsection{Electronic Structure: SDW or CDW instability}

\begin{figure}
\includegraphics[scale=1.0]{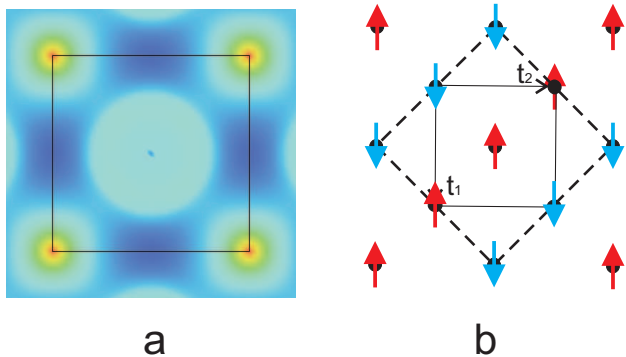}
\caption{Calculated $Re\,\chi(\mathbf{q},0)$ mapped onto the plane through $\Gamma$ with z axis as its normal. The calculated peak value at $M$ is $1.14\times10^{11}$ with the minimum scaled to zero. The right figure shows the SDW derived from $\mathbf{Q}_{M}=(1/2,1/2,0)$, by assuming a linear polarization.} \label{Fig4}
\end{figure}

The above discussions have already shown that the electronic structure of NaFeAs shares many universal features with the Fe-based superconductors. To further investigate the instability of the electronic system of the paramagnetic state, we have also calculated the generalized susceptibility $\chi(\mathbf{q},\omega=0)$ which is defined as follows,
\begin{equation}
\chi(\mathbf{q},\omega)=\sum_{\mathbf{k},j,j'}\dfrac{f(E_{\mathbf{k},j})
-f(E_{\mathbf{k}+\mathbf{q},j'})}{E_{\mathbf{k}+\mathbf{q},j'}-E_{\mathbf{k}j}-\omega+i 0^{+}}.\label{eq:chi}
\end{equation}
The integrand of Eq. \ref{eq:chi} is a integral kernel for constructing a $\mathbf{T}$ matrix which is directly connected to the dielectric constant including local field corrections.\cite{r27} 
\paragraph*{}
To investigate the singularity of $\chi(\mathbf{q},0)$, its values evaluated within the aforementioned dense $\mathbf{q}$-mesh are mapped onto different planes perpendicular to the z axis. Our investigations indicate that the peak of $Re\,\chi(\mathbf{q},0)$ occurs for (1/2,1/2,\textit{z}), but becomes increasingly weak from \textit{z}=0 to the boundary, \textit{z}=1/2. This finding together with the results for $Im\chi(\mathbf{q},0)$ agree quite well with the earlier calculations for the (1:1:1:1) system \cite{r7}. Fig. \ref{Fig4}a 
shows the function values of $Re\,\chi(\mathbf{q},0)$ on the $\Gamma$ plane. It is obvious that a peak appears at $\mathbf{q}$, (1/2,1/2,0), which implies that the condition for a SDW, $\overline{V}_{q}\geq 1/\chi_{q}$ can easily be satisfied at low temperature, \cite{r28,r29} with $\overline{V}_{q}$ being the average exchange matrix element. By assuming a linear SDW with $p(\mathbf{R})=\mathit{p}\,\widehat{\mathbf{x}}\cos(\mathbf{q} \cdot \mathbf{R})$, one attains a spin structure as shown in Fig. \ref{Fig4}b, where \textit{p} has the dimension of spin per unit volume, $\mathbf{R}$ is the lattice position for Fe. The SDW as shown in Fig.  \ref{Fig4}b breaks the original symmetry and supercell of $\sqrt{2}\times \sqrt{2}$ sets in. This type of SDW state has been shown by Yildirim \cite{r11} with DFT calculations to have the lowest energy among the investigated states for LaFeAsO. It is obvious that the nearest-neighbor spin coupling along the two diagonal directions is ferromagnetic and antiferromagnetic, respectively. This asymmetry may favor the tetragonal-orthorhombic transition as observed in other systems. It should to be pointed out that the spin dynamics need to be considered in a more strict way since the \textit{d}-electrons around the Fermi level are itinerant.         
\paragraph*{}
The peak of $Re\,\chi(\mathbf{q},0)$ only implies the instability of an electronic system, which may result in a SDW, or other new orders, in particular a CDW with lattice distortion as discussed by Chan et. al. based on a Hartree-Fock approximation (HFA) with screened exchange interaction. \cite{r29}  The condition for the occurrence of a CDW within the restricted HFA can be written as
\begin{equation}
 \dfrac{4\overline{\eta}_{\mathbf{q}}^{2}}{\hbar \omega _{\mathbf{q}}}-(2\overline{U}_{\mathbf{q}}-\overline{V}_{\mathbf{q}})\geq \dfrac{1}{\chi _{\mathbf{q}}} , \label{eq:CDW} 
\end{equation} 
where $\overline{U}_{\mathbf{q}}$ and $\overline{\eta}_{\mathbf{q}}$ are the average Coulomb interaction and e-p interaction matrix elements, respectively. If the values of $\overline{U}_{\mathbf{q}}$ and  $\overline{V}_{\mathbf{q}}$ are estimated to be of the order of  $\sim$ 4 eV, and $\sim$ 0.7 eV as obtained for LaFeAsO \cite{r24,r30}, then a strong e-p interaction and a soft phonon mode for the nesting $\mathbf{q}$ vector are prerequisites for the CDW to take place. Earlier calculations \cite{r12,r13} using other methods have already revealed a rather weak average e-p coupling, however, no discussions have been made in this respect.
\begin{figure}
\includegraphics[scale=0.9]{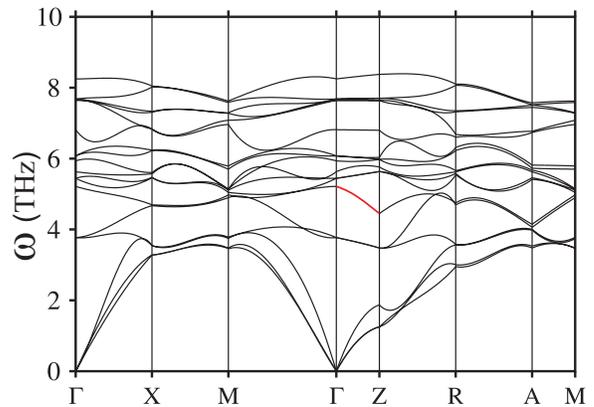} 
\caption{Phonon dispersion calculated from linear-response theory implemented in the FP-LMTO method.} \label{Fig5}
\end{figure} 
\subsection{Lattice dynamics and e-p coupling}

 The calculated phonon spectrum is shown in Fig. \ref{Fig5}, where overall agreement with results using ultrasoft pseudopotential method\cite{r13} is attained. The main differences occur along $\Gamma$-Z, where obviously the authors 
of Ref. \cite{r13} have chosen a different coordinate for the Z point. In our calculations, we have chosen a Z point which lies at the boundary of the standard BZ as shown in Fig. \ref{Fig3}, which agrees also with that of Ref. \cite{r12}. 
\begin{table*}
\caption{The calculated phonon frequencies in $cm^{-1}$ at $\Gamma$ for NaFeAs, the Raman and infrared-(IR) modes are indicated with their irreducible representations in $D_{4h}$. For comparison the values from Ref. \cite{r13} are listed in the last two rows.}\label{tbl2}
\begin{ruledtabular}
\begin{tabular}{ccccccc}
Raman &125.77-$E_{g}$&173.98-$A_{1g}$&198.01-$A_{1g}$&202.65-$E_{g}$&227.35-$B_{1g}$&256.27-$E_{g}$ \\
IR &181.56-$E_{u}$&187.80-$A_{2u}$&254.99-$E_{u}$&275.15-$A_{2u}$\\
Raman &110-$E_{g}$&176-$A_{1g}$&199-$A_{1g}$&187-$E_{g}$&218-$B_{1g}$&241-$E_{g}$ \\
IR &170-$E_{u}$&183-$A_{2u}$&233-$E_{u}$&253-$A_{2u}$\\
\end{tabular}
\end{ruledtabular}
\end{table*}
The detailed comparison for the phonon frequencies at $\Gamma$ is given in Table \ref{tbl2} by transforming the units from THz in Fig. \ref{Fig5} to $cm^{-1}$. The overall agreement between two different calculations is rather good. The largest differences are observed for the two highest IR-active modes, which are only 22.15 and 21.99 $cm^{-1}$ for the $A_{2u}$ and $E_{u}$ modes, respectively. These values can easily be tested experimentally. As shown in Fig. \ref{Fig5}, the phonon spectrum at M does not show any obvious softening. The evaluation of the first term in Eq. \ref{eq:CDW} needs more effort. In the original derivation of Eq. \ref{eq:CDW} unit cells with one more atom were not considered. All e-p matrix elements $\eta (\mathbf{k},\mathbf{k+q},\mathbf{q})$ were treated as an average constant $\overline{\eta}_{\mathbf{q}}$. As a rough estimation, if the actual e-p matrix elements, $\eta (\mathbf{k}j,\mathbf{k+q}j',\mathbf{q} \nu)$ can be replaced by an average, then the condition given in Eq. \ref{eq:CDW} can be used to judge the occurrence of a CDW related to a phonon $\mathbf{q}\nu $. The e-p matrix elements $\eta (\mathbf{k}j,\mathbf{k+q}j',\mathbf{q}\nu)$ can be calculated by linear-responce theory as in Ref. \cite{r15}, where it is symbolized as $g_{\mathbf{k+q}j',\mathbf{k}j}^{\mathbf{q}\nu}$. In this work, we use the following Fermi surface average to estimate $\overline{\eta}_{\mathbf{q}\nu}^{2}$.
\begin{equation}
    \overline{\eta}_{\mathbf{q} \nu}^{2}=\dfrac{1}{N(0)}\sum_{\mathbf{k}jj'} |\eta_{\mathbf{k+q}j',\mathbf{k}j}^{\mathbf{q}\nu} |^2 \delta(\varepsilon_{\mathbf{k}j})\delta(\varepsilon_{\mathbf{k+q}j'}) \label{eq:aveta}
\end{equation}
where $N(0)$ is the density of states at the Fermi level. By using                    
Eq. \ref{eq:aveta}, the first term in Eq. \ref{eq:CDW} can easily be calculated using the phonon spectrum displayed in Fig. \ref{Fig5}. We have calculated $\frac{4\overline{\eta}_{\mathbf{q}\nu}^{2}}{\hbar \omega _{\mathbf{q}\nu}}$ for all other modes except the three trivial acoustic phonon modes. At $\mathbf{q}$,$(0.5,0.5,0)$, the largest value of $\frac{4\overline{\eta}_{\mathbf{q}\nu}^{2}}{\hbar \omega _{\mathbf{q}\nu}}$ of the 18 phonon branches is 0.542 eV with $\nu=3$, which is too small to satisfy Eq. \ref{eq:CDW}. This value is smaller than that of Cr as estimated from the experimental data. \cite{r29} In fact, even when we take the sum of all $|\eta_{\mathbf{k+q}j',\mathbf{k}j}^{\mathbf{q}\nu} |^2$ values at the Fermi surface without averaging, the largest value is 2.08 eV with $\nu=3$, which is still too small to satisfy Eq. \ref{eq:CDW}. We thus exclude the possibility of a CDW with lattice distortion related to the nesting vector. The largest two values on our $\mathbf{q}$ mesh for $\frac{4\overline{\eta}_{\mathbf{q}\nu}^{2}}{\hbar \omega _{\mathbf{q}\nu}}$ are 2.99 and 2.73 eV at $\mathbf{Z}$, (0,0,1/2) and $\mathbf{q}$, (0,0,1/4), respectively, with $\nu=6$, which are still too small to satisfy the criterion of Eq. \ref{eq:CDW}. This fact also implies that the largest e-p interaction occurs in the interval between $\Gamma$ and Z for phonon branch 6 as indicated in Fig. \ref{Fig5}. To demonstrate this result more clearly, we have calculated the phonon mode dependent e-p coupling constant defined as follows,
\begin{equation}
 \lambda_{\mathbf{q} \nu}=\dfrac{1}{\pi N(0)}\dfrac{\gamma_{\mathbf{q}\nu}}{\omega_{\mathbf{q}\nu}^{2}}, \label{eq:lamdaq}
\end{equation} 
where $\gamma_{\mathbf{q}\nu}$ is the phonon line-width as defined in Ref. \cite{r15}. The values $\lambda_{\mathbf{q} \nu}$ for the above two phonon modes are 1.495 and 1.365, respectively, which are also the largest values among all the phonon modes we studied. This consistence indicates that under the approximation of Eq. \ref{eq:aveta}, the first term in Eq. \ref{eq:CDW} is essentially another measure of the e-p coupling. 
\begin{figure}
\includegraphics[scale=1.0]{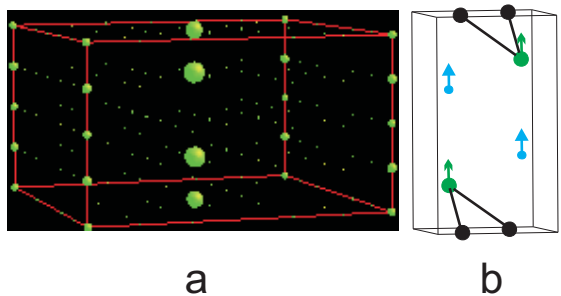}
\caption{$\mathbf{q}$ dependence of e-p coupling constant, $\lambda(\mathbf{q})$, in the first BZ of phonons revealing a peak-like structure; The right figure shows the most important $A_{1g}$ phonon in e-p coupling.} \label{Fig6}
\end{figure} 
To reveal the $\mathbf{q}$ dependence of $\lambda$, we have further summed out the phonon branch index $\nu$ of $\lambda_{\mathbf{q} \nu}$, see Fig. \ref{Fig6}a. As shown in this figure, the characteristic peak-like structure of e-p coupling in superconductors \cite{r31,r32,r33} exists also for NaFeAs. However, in the case of NaFeAs, the maximum value is 3.19 at $\mathbf{q}$, (0,0,1/4), compared to 12.3 in Hg, \cite{r31}, 25.3 in MgB$_{2}$ \cite{r33} and 25 in Li$_{1-x}$BC \cite{r32}. As also shown in Fig. \ref{Fig6}a, the most effective e-p coupling occurs along $\Gamma$-Z. Except for the two largest values, the other effective ones are all at least 4 times smaller, while all remaining ones vanish.  This result is different from those calculated for LaFeAsO$_{1-x}$F$_{x}$, where the e-p coupling distributes relatively evenly among the $\mathbf{q}$ points and phonon branches. \cite{r12} In particular, the most important e-p coupling is at $\Gamma$, where we found no coupling at all. The reason for this difference needs to be clarified in future work since there are many similarities between the two systems. An unusual result is that the dominating e-p coupling stems from phonons involving only As and Na vibrations perpendicular to the Fe-As layer. In Fig. \ref{Fig6}b., the $A_{1g}$ phonon at Z which exhibits the largest $\lambda_{q \nu}$ value is shown. The less effective phonons have      
$\lambda_{q \nu}$ values below 0.3, involving the Fe vibrations in the X-Y plane. However, even for these phonons the dominating component is still based on As, Na vibrations along the z direction. This result indicates that only very few electronic states around the Fermi level are effectively coupled to the phonons. As can be seen from the projected DOSs shown in Fig.  \ref{Fig1}, the weights of As and Na states at the Fermi level are rather small. By calculating the compositions of the Fermi states, we have found that the As and Na show significant presence only in the steep band crossing the $\Gamma-Z$ line as shown in Fig. \ref{Fig1}. A representative state $\psi_{Z,12}$ as indicated in Fig. \ref{Fig1} is plotted in Fig.
\begin{figure}
 \includegraphics[scale=0.8]{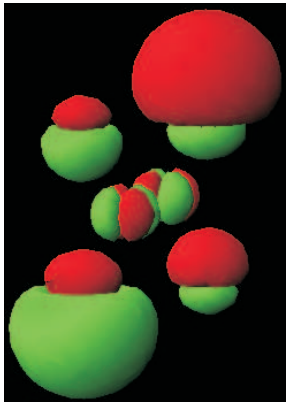}
\caption{A representative state for the steep band which couples effectively with the $A_{1g}$ phonon. The largest lobes indicate the Na-$s,p$ hybrid, the middle sized lobes indicate the As-$s,p$ hybrid, the most contracted lobes indicate the hybrid of mainly Fe-$d_{x^{2}-y^{2}}$ character.} \label{Fig7}
\end{figure} 
\ref{Fig7}. It is understandable that the $d_{x^{2}-y^{2}}$ orbital of Fe is less affected by the vibrations of As and Na due to the small overlap between the relevant orbitals. The large lobe of the Na-$s,p$ hybrid is due to its diffuse character, which is often treated as interstitial electrons in other methods. The weak e-p coupling for specific phonons as discussed above and in particular the rather small percentage of effectively coupled electronic and phononic states in the phase space result in a small total $\lambda$=0.194 compared to 0.27 from the pseudo-potential calculations \cite{r13} and 0.21 for LaFeAsO$_{1-x}$F$_{x}$ \cite{r12}. With this value and the calculated logarithmic average phonon frequency, $\omega_{log}$=215.8K, the calculated superconducting transition temperature $T_{c}$ based on the Allen and Dynes formula $T_{c}$ is close to zero. Even if the multiband effect is considered, the experimental $T_{c}$ of 9K is difficult to be explained within the conventional phonon mechanism. Nevertheless, the discovered dominance of the $A_{1g}$ phonon mode in the e-p coupling may provide a clue to explain the large change of $T_{c}$ when the rare-earth element is substituted by another one.\cite{r35} A possible role played by this type of phonons is the magnetoelastic effect, since it is found that the magnetic coupling depends strongly on the z coordinate of the As atom.\cite{r10} It needs to be pointed out that after we have finished our calculations we were informed by Egami that in BaFe$_{2}$As$_{2}$ the most important phonon mode is indeed of $A_{1g}$ type exhibiting significant softening as a result of inelastic X-ray scattering measurements.\cite{r36}.
\paragraph*{}
The dependence of e-p coupling on electron or hole doping needs to be investigated in future work, because the relative positions of the $E_{g}$, $B_{1g}$ flat bands with respect to the Fermi level change with doping. Therefore, the structure and strength of e-p coupling may change. Though these states dominate at the Fermi level as revealed in this work and other calculations, they have negligible e-p coupling in the present calculations. We have already shown that the Coulomb interaction is a prohibiting factor for a CDW, however, the correlation effect needs further discussion. It has already been argued that the Fe pnictides are also in proximity to Mott insulators. \cite{r24,r37} By using the calculated potential parameter, we have estimated the band width W$_{3d}$=3.5 eV for the Fe-$3d$ orbital as compared to the Coulomb parameter $U\sim$ 4eV. This implies that NaFeAs should be a Mott insulator or more accurately a bad metal considering the degeneracy of d bands.\cite{r24} When correlation effects are considered at the DMFT level \cite{r24}, very large spectral weight transfers from $B_{1g}-d_{x^{2}-y^{2}}$ to the $A_{1g}-d_{3z^{2}-1}$ and a small increase for the As-$4p$ orbital occur. Considering the established similarity between the two systems, we would expect an analogous transfer to take place, which means the $d_{x^{2}-y^{2}}$ component in Fig. \ref{Fig7} will be replaced by $ d_{3z^{2}-1}$ and thus increase the Fe-As interaction along the z-direction. Accordingly, the e-p coupling between this orbital and the $A_{1g}$ mode would be expected to increase. This needs to be confirmed the future.

\section{CONCLUSION}
Based on first-principles calculations, we have shown the similarity between the electronic structures of NaFeAs and other Fe pnictides. By using a tight-binding model, a global double degeneracy of bands along X-M and R-A directions has been revealed caused by the combination of site symmetry of Fe, As and the line symmetry of X-M in the BZ. The studies on the topology, metric property of FS and the generalized susceptibility $\chi(\mathbf{q})$ demonstrate the paramagnetic state of NaFeAs to be unstable against the $\mathbf{Q}_{M}$=(1/2,1/2,0) SDW state. Thus a SDW ground state with a possible structural distortion accompanied with symmetry breaking are predicted, a fact confirmed by recent experiments. \cite{r38} The possibility of a $\mathbf{Q}_{M}$ related CDW can be completely excluded based on a resticted HFA criterion. A $A_{1g}$ phonon involving only the z-direction vibration of As and Na has been detected. A possible enhancement of this specific e-p coupling due to the correlation effect has also been discussed.


\begin{thebibliography}{99}

\bibitem{r1} Y. Kamihara, T. Watanabe, M. Hirano, and H. Hosono, J. Am. Chem. Soc. {\bf 130},3296 (2008).
\bibitem{r2} J. W. Lynn and P. Dai, Physica C {469}, 469 (2009).
\bibitem{r3} M. J. Pitcher, D. R. Parker, P. Adamson, S. J. C. Herkelrath,
A. T. Boothroyd, and S. J. Clarke, Chem. Commun.,
5918 (2008).
\bibitem{r4} J. H. Tapp, Z. Tang, B. Lv, K. Sasmal, B. Lorenz, P. C. W.
Chu, and A. M. Guloy, Phys. Rev. B {\bf 78}, 060505 (2008).
\bibitem{r5} D. R. Parker, M. J. Pitcher, and S. J. Clarke, Chem. Comm. {\bf 16}, 2189 (2009).
\bibitem{r6} X. C. Wang, Q. Q. Liu, Y. X. Lv, W. B. Gao, L. X. Yang,
R. C. Yu, F. Y. Li, and C. Q. Jin, Solid State Comm. {\bf 148}, 538 (2008).
\bibitem{r7} I. I. Mazin, D. J. Singh, M. D. Johannes, and M. H. Du, Phys.
Rev. Lett. {\bf 101}, 057003 (2008).
\bibitem{r8} J. Dong, H. J. Zhang, G. Xu, Z. Li, G. Li, W. Z. Hu, D.
Wu, G. F. Chen, X. Dai, J. L. Luo, Z. Fang, and N. L.
Wang, Europhys. Lett. {\bf 83}, 27006 (2008).
\bibitem{r9} D. J. Singh, Phys. Rev. B {\bf 78}, 094511 (2008).
\bibitem{r10} Z. P. Yin, S. Leb\`{e}gue, M. J. Han, B. Neal, S. Y. Savrasov,
and W. E. Pickett, Phys. Rev. Lett. {\bf 101}, 047001 (2008).
\bibitem{r11} T. Yildirim, Phys. Rev. Lett. {\bf 101}, 057010 (2008).
\bibitem{r12} L. Boeri, O. V. Dolgov, and A. A. Golubov, Phys. Rev.
Lett. {\bf 101}, 026403 (2008).
\bibitem{r13} R. A. Jishi and H. M. Alyahyaei, Phys. Rev. B {\bf 79}, 064516 (2009).
\bibitem{r14} S. Y. Savrasov, Phys. Rev. B {\bf 54} 16470, (1996).
\bibitem{r15} S. Y. Savrasov and D. Y. Savrasov, Phys. Rev. B {\bf 54} 16487, (1996).
\bibitem{r16} J. F. Janak, V. L. Moruzzi, and A. R. Williams,Phys. Rev. B {\bf 12} 1257, (1975).
\bibitem{r17} J. P. Perdew and Y. Wang, Phys. Rev. B {\bf 45} 13244, (1992).
\bibitem{r18} O. K. Andersen, Phys. Rev. B {\bf 12} 3060, (1975).
\bibitem{r19} O. K. Andersen, and O. Jepsen, Phys. Rev. Lett. {\bf 53} 2571, (1984).
\bibitem{r20} I. A. Nekrasov, Z. V. Pchelkina, and M. V. Sadovskii,
JETP Lett. {\bf 88} 543, (2008).
\bibitem{r21} I. A. Nekrasov, Z. V. Pchelkina, and M. V. Sadovskii,
JETP Lett. {\bf 88} 144, (2008).
\bibitem{r22} N. W. Ashcroft and N. D. Mermin, \emph{Solid State Physics}, (Saunders,Philadelphia, 1976, pp. 107-108).
\bibitem{r23} R. Hoffmann, Angew. Chem. Int. Ed. Engl. {\bf 26} 846, (1987).
\bibitem{r24} K. Haule, J. H. Shim, and G. Kotliar, Phys. Rev. Lett.
{\bf 100}, 226402 (2008).
\bibitem{r25} J. M. An, and W. E. Picket, Phys. Rev. Lett.
{\bf 86}, 4366 (2001).
\bibitem{r26} S. Deng, A. Simon, and J. K\"ohler, J. Supercond. 
{\bf 16}, 477 (2003).
\bibitem{r27} N. Wiser, Phys. Rev. {\bf 129}, 62 (1963).
\bibitem{r28} P.A. Fedders and P. C. Martin, Phys. Rev. {\bf 143}, 245 (1967).
\bibitem{r29} S. -K Chan and V. Heine, J. Phys. F. {\bf 3}, 795 (1973).
\bibitem{r30} T. Miyake and F. Aryasetiawan, Phys. Rev. B {\bf 77} 085122, (2008).
\bibitem{r31} S. Deng, A. Simon, and J. K\"{o}hler, J. Phys. Chem. Solids {\bf 62} 1441, (2001).
\bibitem{r32} J. M. An, S. Y. Savrasov, H. Rosner, and W. E., Pickett, Phys. Rev. B {\bf 66} 220502(R), (2002).
\bibitem{r33} S. Deng, A. Simon, and J. K\"{o}hler, J. Supercond. {\bf 16}, 477, (2003).
\bibitem{r34} P. B. Allen and R. C. Dynes, Phys. Rev. B {\bf 12}, 905 (1975).
\bibitem{r35} Z. -A. Ren, W. Lu, J. Yang, W. Yi, X.-L Shen, Z. -C Li, G. -C. Che, X. -L. Dong, L. -L. Sun, F. Zhou, and Z. -X. Zhao, Chin. Phys. Lett. {\bf 25}, 2215 (2008).
\bibitem{r36} D. Reznik, K. Lokshin, D. C. Mitchell, D. Parshall, W. Dmowski, D. Lamago, R. Heid,
K.-P- Bohnen, A. S. Sefat, M. A. McGuire, B. C. Sales, D. G. Mandrus, A.  Asubedi, D. J. Singh, A. Alatas, M. H. Upton, A. H. Said, Yu. Shvyd’ko, and T. Egami, arXiv : {\bf 0810.4941} (2008).
\bibitem{r37} Q. Si and E. Abrahams, Phys. Rev. Lett. {\bf 101}, 076401 (2008).
\bibitem{r38} S. Li, C. de la Cruz, Q. Huang, G. F. Chen, T. -L. Xia, J. L. Lou, N. L. Wang, and P. Dai, Phys. Rev. B {\bf 80}, 020504 (2009). 
\end{thebibliography}
\end{document}